\documentclass{article}
\usepackage{spconf,amsmath,graphicx}
\usepackage{setspace}
\usepackage{enumitem}
\usepackage{url}
\usepackage[noend]{algpseudocode}
\usepackage{algorithmicx,algorithm}
\usepackage{makecell}
\usepackage{multirow}

\title{Attack on Practical Speaker Verification System using universal adversarial perturbations}
\name{\begin{tabular}{c}Weiyi Zhang$^1$, Shuning Zhao$^1$, Le Liu$^3$, Jianmin Li$^1$ \\
Xingliang Cheng$^2$, Thomas Fang Zheng$^2$, Xiaolin Hu$^{{\star}1}$\thanks{${\star}$Corresponding author}\end{tabular}}

\address{$^{1}$ Department of Computer Science and Technology, BNRist, Tsinghua University, China\\
      $^2$ Center for Speech and Language Technologies, BNRist, Tsinghua University, China\\
      $^3$ Beijing d-Ear Technologies Co.,Ltd. }

\begin{document}


\maketitle
\begin{abstract}
In authentication scenarios, applications of practical speaker verification systems usually require a person to read a dynamic authentication text. Previous studies played an audio adversarial example as a digital signal to perform physical attacks, which would be easily rejected by audio replay detection modules. This work shows that by playing our crafted adversarial perturbation as a separate source when the adversary is speaking, the practical speaker verification system will misjudge the adversary as a target speaker. A two-step algorithm is proposed to optimize the universal adversarial perturbation to be text-independent and has little effect on the authentication text recognition. We also estimated room impulse response (RIR) in the algorithm which allowed the perturbation to be effective after being played over the air. In the physical experiment, we achieved targeted attacks with success rate of 100\%, while the word error rate (WER) on speech recognition was only increased by 3.55\%. And recorded audios could pass replay detection for the live person speaking.

\end{abstract}
\section{Introduction}
\label{sec:intro}
The automatic speaker verification (ASV) process is a convenient and reliable process for identity verification. Many authentication scenarios \cite{reynolds2002overview} such as device access control, banking activities and forensics use ASV for verification. DNN-based ASV models \cite{snyder2018x, xie2019utterance, chung2020defence} tend to have excellent performance, but many studies have shown that audio adversarial examples can make the ASV process give wrong decisions \cite{li2020adversarial, liu2019adversarial} or let adversary pass verification \cite{chen2019real, li2020learning}. The transferability of audio adversarial examples across different models was also revealed in \cite{li2020adversarial, liu2019adversarial}. Audio adversarial examples could still remain effective after being played over the air in \cite{li2020practical}.

\begin{figure}[ht]
\centering
\includegraphics[scale=0.5]{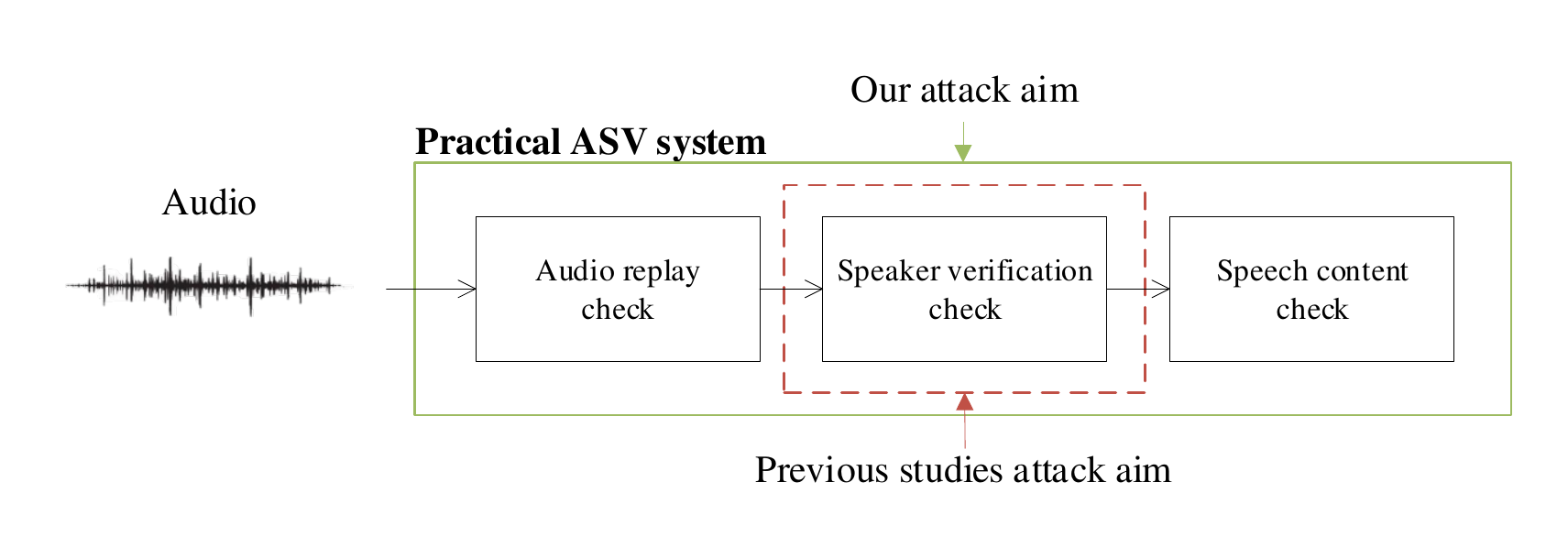}
\caption{Schematic diagram of attacking the PSV system}
\label{fig:1}
\end{figure}

In real applications (e.g., Alipay APP \cite{alipay} and China Construction Bank APP \cite{ccb}), when a user starts the ASV process, unfixed texts (e.g., random numbers) are sent to the user from the server. After the user speaking the same content speech, the audio recorded by a microphone will go through three-module checks: audio replay check, speaker identity check, and speech content check. Only when all three check parts give pass decision, the user can be verified successfully as shown in Figure \ref{fig:1}. We call it the practical speaker verification (PSV) system. Previous studies \cite{li2020adversarial, liu2019adversarial, chen2019real, li2020learning, li2020practical} only consider attacking the speaker identity check module to let it break. But their adversarial examples will be rejected in the PSV system for audio replay or different speech content. Studies \cite{li2020universal, xie2020real} crafted universal adversarial perturbations that were text-independent and could launch attack in real time. But it is not proved that their perturbations could not affect the speech content recognition and remained effective after being played separately over the air to pass the audio replay detection.

In this paper, we propose a two-step algorithm to craft universal adversarial perturbations adapted to attack the PSV system. We combine Carlini-Wagner (CW) objective function \cite{carlini2017towards} and Projected Gradient Descent (PGD) \cite{madry2017towards} methods to perform targeted attack on a DNN-based speaker verification model \cite{chung2020defence}. Three properties of our adversarial perturbations can be concluded as follow: 

\begin{itemize}[leftmargin = 15pt, nosep]
\item \textbf{Targeted} \,We craft an adversarial perturbation that misleads the ASV model to verify the identity of the adversary’s audio as a targeted speaker. 
\item \textbf{Universal} \,The adversarial perturbation is text independent. It is effective whatever content the adversary speaks and has little effect on the speech content recognition.
\item \textbf{Robust} \,Incorporating RIR into the generation process of perturbation help it be effective after being played over the air. When the adversary is speaking to perform a physical attack, the adversarial perturbation is played as separated source, which helps to pass the audio replay detection.
\end{itemize}

We achieved successful targeted attack both in digital and physical experiments. Especially in physical experiments with 3 volunteers, with a 100\% attack success rate, our adversarial attack could pass audio replay detection and increase speech recognition WER only by 3.55\%.

\section{Practical Speaker verification}
\label{sec:psv}

As we demonstrate above, the practical speaker verification system is commonly used in real applications and includes three-module checks: audio replay check, speaker identity check, and speech content check as shown in Figure \ref{fig:1}. We will describe our three-module checks separately.

\noindent\textit{(1) Audio replay check.} \,We use the audio replay detection model in \cite{cheng2019replay} to perform audio replay check, which won the first place in the physical access of ASVspoof 2019.

\noindent\textit{(2) Speech content check.} \,We use the google cloud speech to text API \cite{GCP} as the speech recognition model for speech content check. Word error rate (WER) and character error rate (CER) are the metric to measure the level of distortion to the speech contents caused by the adversarial perturbations.

\noindent\textit{(3) Speaker identity check.} \,This check mainly calculates the distance between the speaker feature representation of new input audio and that of the enrolled audio. We use a DNN-based speaker verification model in \cite{chung2020defence} to encoder the speaker feature representation from audio. The model first creates Mel-spectrogram for a raw audio signal. And the following are ResNet-34 convolutional layers to extract frame-level features. Then attentive statistics pooling is used to aggregate frame-level features into utterance-level representation, which produces both means and standard deviations with importance weighting utilizing attention. Finally, a fully connected layer is used to map the statistics into a fixed dimension vector witch is the speaker feature representation. More details about the ASV model can be found in \cite{chung2020defence}.

Let $F(x)$ denotes the DNN-based speaker verification model \cite{chung2020defence}, which receives an input audio $x$ and gives the speaker feature representation $v=F(x)$. For two audios $x_1$ and $x_2$, we use cosine similarity as a score to measure the distance between their speaker feature representations. The score $s(F(x_1), F(x_2))$ is calculated by 
\begin{equation}
    \label{eq:cos_sim}
    s(F(x_1), F(x_2)) = \frac{F(x_1) F(x_2)}{\left\|F(x_1)\right\|_2\left\|F(x_2)\right\|_2}.
\end{equation}
It will give a same speaker decision when the score satisfies $s(F(x_1), F(x_2))\geq\theta$, where $\theta$ is a preset threshold. 

We aim to attack the PSV syetem shown in Figure \ref{fig:1}. It means that we only do an adversarial attack on the speaker verification model but the audio replay check and speech content check should not fail with our adversarial perturbations. The adversarial perturbation can lead the speaker verification model to verify an adversary as the enrolled targeted speaker. When the required speech content changes, it can be applied directly and needn't be crafted again.

\section{Targeted, universal, and robust adversarial perturbations}
\label{method}

\subsection{Attack on the ASV model}
\label{sec:attack}

We consider a white-box threat model where the adversary has full knowledge about the ASV model but no knowledge about the audio replay detection and speech recognition model. The generation of our adversarial perturbations is described as follow. 

There is a targeted speaker with the enrolled audio $y$. The content of an adversary's audio $x$ is specific to text $t$. Our adversarial perturbation $\delta$ has a fixed length. To launch an adversarial attack, we need to repeat $\delta$ to get $\delta^{'}$ which has a same length as the input audio $x$. Our aim is to find an adversarial perturbation $\delta$ such that (a) $s(F(x+\delta^{'}),F(y)) \geq \theta$, (b) $\delta$ is text-independent and (c) the speech recognition result of audio $x+\delta^{'}$ is $t$. We propose a two-step algorithm to optimize the adversarial perturbation.

In the first step, we maximize the attack effect on the ASV model which is similar to the method in \cite{xie2020real}. We make $\delta$ be effective to lead a targeted attack on the ASV model regardless of the content of input $x$. $N$ audios of the adversary are collected to form a training set $X=\{x_1, x_2, ... ,x_N\}$ where each $x_i$ contains different text contents. If $N$ is large enough, the training set $X$ will cover great diversity about the adversary such as start offset, tune, emotion, speech content and etc. Training on $X$ can make $\delta$ easily transferable to other new audios of the same adversary. We define a CW-like function as the minimization objective:
\begin{equation}
    \label{eq:loss1}
    L_1(X, \delta) = \sum_{i=1}^{N}{\rm max}(\theta - s(F(x_i+\delta^{'}),F(y)), -\kappa)
\end{equation}
where $s(F(x_i+\delta^{'}),F(y))$ is the score between the feature representation of the adversary and targeted speaker. $\kappa$ is the attack confidence such that a large $\kappa$ can get high attack success rate on other test audios of the adversary.

The $l$ is the max audio length in $X$. We repeat $\delta$ and each train audio until its length is $l$. The $\delta^{'}$ is added into each train audio to form a $N$-batch data so that $L_1(X, \delta)$ can be calculated in one forward propagation. To minimize $L_1(X, \delta)$ we use $l_{\infty}$ PGD with momentum method to update $\delta$. The update rule for $\delta$ is
\begin{equation}
    \label{eq:update1}
    \begin{split}
        &g = \beta g_{i-1} + g_i\\
        &\delta_{i+1} = Clip_{\epsilon}(\delta_i - \alpha_1 sign(g))
    \end{split}
\end{equation}
where the gradient $g_{i} = \partial L_1 / \partial \delta_i$ for the $i$-th step and $g_{i-1}$ for the $(i-1)$-th step are obtained by backpropagation. $g$ is the momentum gradient with hyperparameter $\beta$. $\epsilon$ is the attack strength and $\alpha_1$ is the attack step size. $Clip_{\epsilon}(\delta)$ performs element-wise clipping of $\delta$ into the interval $[-\epsilon, \epsilon]$. There are at most M iterations. The best $\delta_{1}^{*}$ generated by the first step can mislead the ASV model to verify other test audios of the adversary as the enrolled targeted speaker. However, a large $\epsilon$ is usually used especially when considering RIR transformation in Section \ref{rir}. So the speech content recognition is greatly affected by $\delta_{1}^{*}$.

\subsection{Correct speech recognition}
\label{sec:correct}

In the second step, our main purpose is optimizing $\delta$ to reduce the impact on speech recognition. The state-of-the-art speech recognition models usually first extract frequency domain features from audio, like Mel-spectrum or Mel-frequency cepstral coefficients. Minimizing the difference between frequency domain features of $x+\delta^{'}$ and $x$ can have two similar text recognition results. We use ${\rm STFT}(x)$ to represent the short-time fourier transform of the input speech $x$. For the linearity of Discrete Fourier Transform (DFT), we have the difference $d(x+\delta^{'}, x)=|{\rm STFT}(x+\delta^{'}) - {\rm STFT}(x)|=|{\rm STFT}(\delta^{'})|$. So we define another objective function:
\begin{equation}
    \label{eq:loss2}
    L_2(X, \delta) = {\rm mean}(|{\rm STFT}(\delta)|)
\end{equation}

Meanwhile, we need to retain adversarial attack on the ASV model. So the goal of the second step is to minimize the funciton
\begin{equation}
    \label{eq:loss}
    L(X, \delta) = L_1(X, \delta) + \gamma L_2(X, \delta)
\end{equation}
where $\gamma$ is a balanced parameter between two terms. We start the second step from initializing $\delta$ as $\delta_{1}^{*}$ in the first step and continue to optimize $\delta$ to get final adversarial perturbation $\delta^{*}$.

\subsection{Physical Robustness}
\label{rir}


There are two challenges in real application: the audio replay detection and the distortion brought by hardware and physical signal path. Different from previous manner where the audio adversarial example is played as a digital signal, we play the adversarial perturbation as a separate source when the adversary is speaking. There is only one play-record process in our attack but two process in previous manner. So it is easier for our adversarial example to pass the audio replay detection.

To model the distortion from speaker playing to microphone recording, we use the SineSweep method in \cite{stan2002comparison} to estimate the RIR in a room. The special signal $x(t)$ is played by the speaker and $y(t)$ is the audio recorded by the microphone:
\begin{equation}
    \label{eq:sinesweep}
    x(t)={\rm sin}\left(\frac{2\pi f_1 T}{{\rm ln}(\frac{f_2}{f_1})}(e^{\frac{t}{T}{\rm ln}(\frac{f_2}{f_1})} - 1)\right) 
\end{equation}
where $f_1$, $f_2$ are the start and stop frequencies that we want to estimate RIR between, T is the signal duration. The RIR $r(t)$ can be estimated by convolving $y(t)$ with the time-reversal of $x(t)$: $r(t)=y(t)*x(-t)$ where * denotes the convolution operation.

Incorporating the RIR $r(t)$ into the generation of $\delta$ by a transform ${\rm T}(x)=x*r$ will reduce the impact of distortion brought by hardware and physical signal path. The improved function $L_1'(X, \delta)$ is:
\begin{equation}
    \label{eq:loss1_rir}
    L_1'(X, \delta) = \sum_{n=1}^{N}{\rm max}(\theta - s(F(T(x_n)+T(\delta^{'})),F(y)), -\kappa)
\end{equation}

\section{EXPERIMENTAL RESULTS}
\label{result}
\subsection{Experimental Methodology}
\textbf{Dataset} \,LibriSpeech\cite{panayotov2015librispeech} test clean dataset was used to evaluate our adversarial perturbations. It consists of 40 speakers: 20 males and 20 females. It also provided us with text reference so that we could measure WER changes caused by our adversarial perturbations.

\noindent\textbf{Model} \,For three modules in the PSV system, the ASV model \cite{chung2020defence} was trained on Voxceleb2 \cite{nagrani2020voxceleb} train dataset with data augmentation method in \cite{ko2017study}. It is optimized by softmax angular prototypical loss function and reach a 1.36\% EER on LibriSpeech test clean dataset. The threshold obtained when calculating EER would be used as the preset threshold $\theta$ of our entire experiment. Audio replay detection model in \cite{cheng2019replay} and speech recognition model in \cite{GCP} were well pretrained  and we only used them to do evaluation.

\noindent\textbf{Attack Settings} \,We defined two types of adversarial attacks: intra-gender and inter-gender. In digital attacks, there were 20 male and 20 female speakers in the evaluation set. Every speaker would be an adversary. We randomly chose one targeted speaker with the same gender and one targeted speaker with a different gender for each adversary. For every speaker, we selected $N=15$ audios for training, and the rest audios for testing. Totally, there were 40(number of speakers) $\times$ 2(intra-gender and inter-gender) $\times$ 2(without RIR and with RIR) $=$ 160 digital attacks. In physical attacks, we performed 4 attacks among volunteers: two males and one female. Two intra-gender attacks meant that two males took turns: one as the adversary and the other as the targeted speaker. Two inter-gender attacks referred that one female was the adversary and two males were the targeted speaker separately.

\subsection{Evaluation of Digital Attacks}

We set $\epsilon=0.03$ for digital attack without RIR. The results are shown in Table \ref{tab1}. The original scenario is the evaluation of the clean audios. It provides a baseline WER to measure the distortion on speech recognition caused by the adversarial perturbations. As mentioned in Section \ref{sec:attack} our first step is very similar to the method used in \cite{xie2020real}. Hence we used the adversarial perturbation $\delta_{1}^{*}$ generated in the first step as the baseline method. Compared to it, our adversarial perturbation leads to lower WER and higher signal-to-noise ratio (SNR) when get similar targeted attack success rates (ASR). It reveals that our second step optimization is helpful for more accurate speech recognition and better audio quality. The results also show that it is more difficult to conduct inter-gender attacks than intra-gender attacks. More average iterations referred to steps in Table \ref{tab1} are needed for the former to achieve similar targeted ASR as the latter. 

\begin{table}[t]
    \small
	\renewcommand{\arraystretch}{1.3}
	\caption{Results of digital attacks}
	\label{tab1}
	\centering
	\setlength{\tabcolsep}{1mm}{
	\begin{tabular}{|c|c|c|c|c|c|}
		\hline
		Scenario  & Method & Steps & ASR(\%) & WER(\%) & SNR(dB)\\
		\hline
		Original & N/A & N/A           & 0  & 12.95  & N/A\\
		\hline
		\multirow{2}*{Intra-gender} & baseline&236           & 98.43  & 32.33  & 16.90\\
		\cline{2-6}
		~ &ours &846           & 98.65  & 19.43  & 23.66\\
		\hline
		\multirow{2}*{Inter-gender} & baseline&617           & 96.63  & 37.57  & 16.55\\
		\cline{2-6}
		~ &ours &1872           & 96.40  & 21.53  & 22.26\\
		\hline
	\end{tabular}}
\end{table}

We also considered the impact of RIR for the digital attacks and set $\epsilon=0.05$. The actual measured RIR dataset in \cite{szoke2019building} was used as the transformation $T(\cdot)$. There were 773 RIRs for training and 242 RIRs for testing. The results for intra- and inter-gender attack with RIR is similar to that without RIR. Only RIR led to a large increase in the WER for all results, which can be found at \url{https://github.com/zhang-wy15/Attack_practical_asv.}

\subsection{Evaluation of Physical Attacks}
\begin{table}[t]
    \small
	\renewcommand{\arraystretch}{1.3}
	\caption{Results of intra-gender physical attack}
	\label{tab3}
	\centering
	\begin{tabular}{|c|c|c|c|c|}
		\hline
		Scenario  & ASR(\%) & WER(\%) & CER(\%)\\
		\hline
		Clean &0& 11.42 & 5.78\\
		\hline
		Gaussian &0& 17.77          & 10.06\\
		\hline
		Baseline & 80&21.82        & 14.48\\
		\hline
		Ours &100& 14.97           & 7.53\\
		\hline
	\end{tabular}
\end{table}

\begin{figure}[t]
\centering
\includegraphics[scale=0.0449]{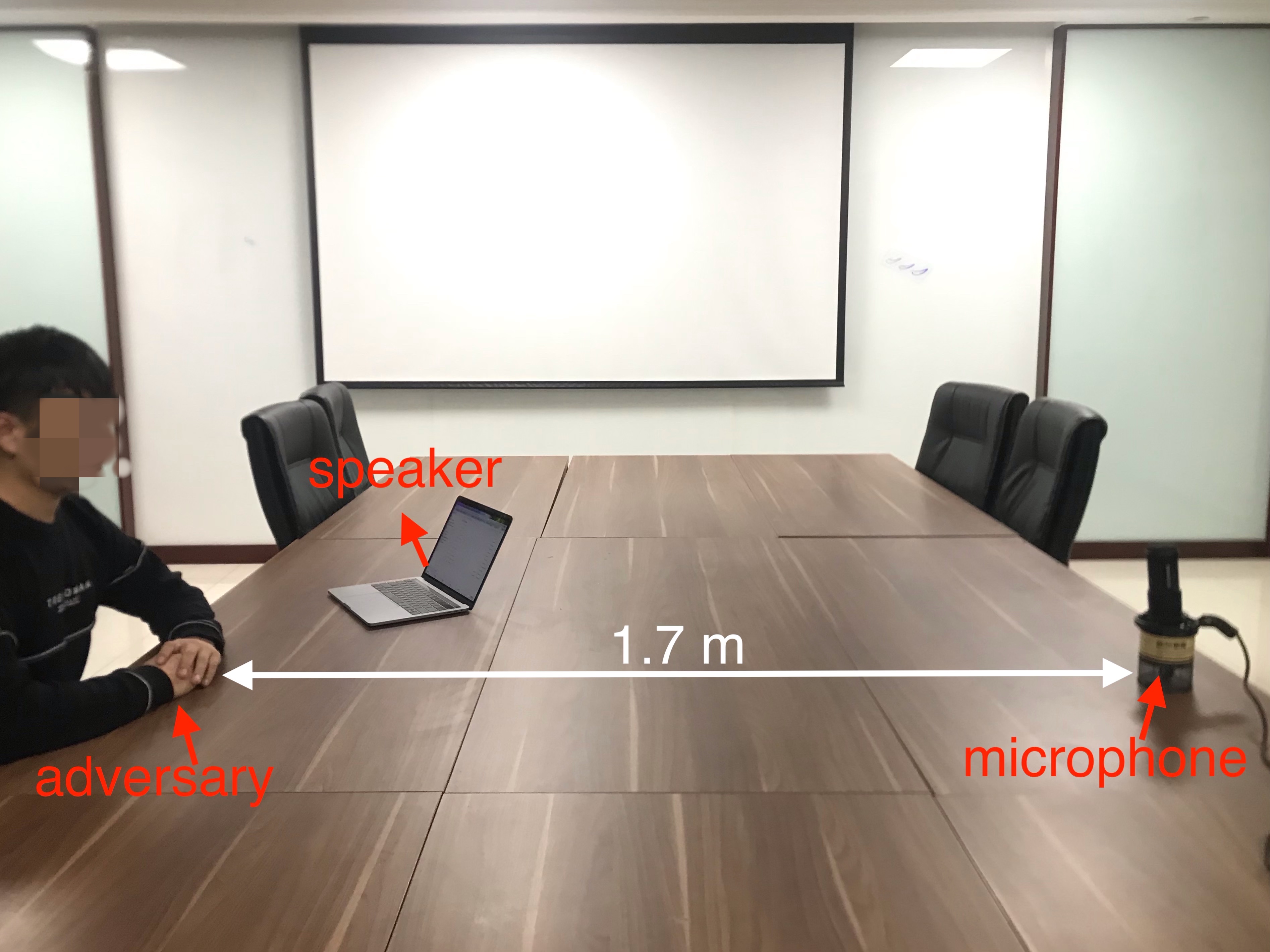}
\caption{Photo of physical attack}
\label{fig:2}
\end{figure}

The perturbation in the physical attack has the same optimization setting as the digital attacks with RIR. We measured real RIRs in a 7.64m width, 8.75m length and 2.4m height meeting room using sine sweept signal. As shown in Figure \ref{fig:2}, the adversarial perturbations were played by the built-in speaker of the MacBook near the adversary. We record the volunteer's speech using a Seeknature T2058 microphone. The distance between adversary and microphone is 1.7m.

We defined 15 sentences that were not previously seen in the training set as our authentication text. To demonstrate the superiority of our adversarial perturbation we used three other methods for comparison. The adversary had to read the same speech contents four times. In the first scenario, we played nothing and only recorded the test speech of adversary as clean benchmark. In the second scenario, the speaker played gaussian noise with equal $l_{\infty}$ norm to our adversarial perturbation when the adversary was speaking. In the two final scenarios, the speaker played the corresponding perturbation for the baseline and our proposed methods. The intra-gender attack results in Table \ref{tab3} show that our adversarial perturbation had a 100\% attack success rate which is 20\% higher than the baseline methods. For speech recognition, our method only increased the WER by 3.55\% compared to the clean speech, while the baseline method increased the WER by 10.40\%. Our perturbation even outperformed the Gaussian noise which proves the effectiveness of our second-step algorithm to reduce the impact on speech recognition.

To illustrate the importance of live human volunteer, we performed audio replay detection using the model in \cite{cheng2019replay}. We collected 45 audio adversarial examples from the previous studies \cite{li2020adversarial, chen2019real, li2020learning} and 120 our physical adversarial examples. When performing physical attack, their adversarial examples had to be played by a speaker device, but our attack can be conducted with a live human adversary. As expected our adversarial examples had a 67.7\% success rate to pass the replay detection, whereas the adversarial examples from previous studies only had a 37.7\% success rate to pass audio replay detection. It reveals that replay adversarial examples are easier to be rejected.

\section{Conclusion}

\label{conclusion}
We proposed a two-step algorithm to generate adversarial perturbation for attacking the practical speaker verification system. Our perturbation is targeted, universal, and physically robust. It can mislead the PSV system to verify the adversary as a targeted victim. The perturbation is text-independent and have little effect on speech recognition in physical environments. It can also be played as a separate source when the adversary is speaking to pass audio replay detection. We study the vulnerability of PSV system in physical world and help researchers to improve the security of such applications.

\noindent\textbf{Acknowledgments}
This work was supported by the National Natural Science Foundation of China under Grant Nos. U19B2034, 61620106010, 61836014.

\clearpage

\bibliographystyle{IEEEbib}
\begin{spacing}{0.8}
\bibliography{strings,refs}
\end{spacing}
\end{document}